\documentclass[12pt]{iopart}
\usepackage{iopams}
\usepackage{graphicx,color,epsfig,rotate}

\newcommand{\ag}{a$_{1g}$ }
\newcommand{\egp}{e$_g^\prime$ }

\begin{document}
\title{Competing electronic states in high temperature phase of NaTiO$_2$}
\author{Monika Dhariwal$^{1}$, L. Pisani$^{2}$ and T.Maitra$^{1,*}$}
\address{$^{1}$Department of Physics,Indian Institute of Technology Roorkee, Roorkee- 247667, Uttarakhand, India}

\address{$^{2}$via Salvo d'Acquisto 6, 62017 Porto Recanati (MC), Italy}
\ead{tulimfph@iitr.ac.in($^{*}$corresponding author)}

\begin{abstract}
First principle density functional theory (DFT) calculations on the high 
temperature phase of layered triangular lattice system NaTiO$_2$ have revealed 
that there exists a collective electronic state energetically close to the 
ground state but with competing transport properties: the latter is metallic 
with partially occupied doubly degenerate \egp orbitals whereas the former is 
insulating with \ag orbital fully occupied. Significant occupation of this 
excited state is possible at non zero temperature either thermally or thanks 
to very soft (large amplitude) oxygen vibrations. Possible explanations of the 
experimental low conductivity based on competing orbital transport 
and of the specific heat jump at a structural transition based on orbital 
entropy are discussed.
\end{abstract}
\vspace{0.5cm}
\maketitle

\section{Introduction}

Frustrated magnetic systems with active orbital degrees of freedom have drawn
a lot of attention recently among the condensed matter community because of 
their potential in applications such as spintronics\cite{spintr}. 
Two classes of systems that are extensively studied in this context 
are magnetic spinels having transition metal ion in a pyrochlore lattice
\cite{garlea,tchernychov,tm-rv,wheeler} 
and layered materials with transition metal ion forming a two 
dimensional triangular/hexagonal lattice\cite{clarke,cava,khomskii}. 
Recent experiments\cite{cava,jinno} and theoretical calculations\cite{dhariwal,
jia} have shown that orbital ordering in these systems play a major role in 
lifting the geometrical frustration (often via a structural transition) and 
driving a (often unconventional) magnetic order as temperature is lowered. At 
high temperatures, when the lattice is completely frustrated there is a 
possibility of having competing electronic states. If such states are 
accessible by tuning an external parameter such as pressure, temperature or 
magnetic field, one can consider using the system as a switch or memory device 
in spintronics/electronics applications.

The high temperature phase of NaTiO$_2$ seems to be a promising candidate
in this context as we infer from our detail electronic structure 
calculations (discussed below). We observe the presence of two distinct 
electronic states in this system;
one is insulating with \ag orbital fully occupied and the other is metallic 
with partially occupied doubly degenerate \egp orbitals and 
these two states are found to be very close by in energy.     

\begin{figure}
\begin{center}
\includegraphics[width=7cm]{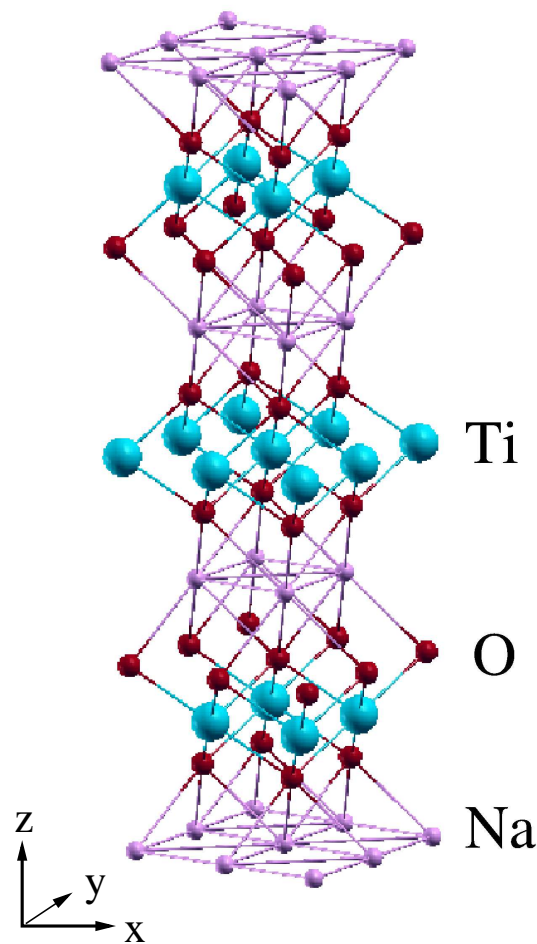}
\caption{Crystal structure of NaTiO$_2$ showing layered triangular lattice 
formed by Ti ions.}
\label{crysstr}
\end{center}
\end{figure}

NaTiO$_2$ in its high temperature phase has rhombohedral symmetry (space group 
R$\bar{3}$m). The two dimensional layer of Ti ions has six equidistant nearest
neighbours having Ti-Ti bond length 3.045 \AA. Each Ti ion is surrounded by
six equidistant O-ions forming TiO$_6$ octahedron with Ti-O bond lengths being
2.078 \AA. Furthermore, these TiO$_6$ octahedra have a trigonal distortion as 
O-Ti-O angles deviate from 90 degrees. 
The three dimensional structure of NaTiO$_2$ is formed by the stacking of 
layers of Na-TiO$_6$-Na. 
These two dimensional layers of Ti ions form a triangular lattice 
in the x-y plane and the z-axis (trigonal axis) passes through the
centroid of O-triangles above and below the Ti layer. 
The x and y-axes point towards the next nearest neighbour Ti
ions and nearest neighbor Ti ions respectively. 
The interlayer Ti-Ti distance is about 6 \AA\  and is much larger compared to
the intralayer Ti-Ti distance. Hence the system can effectively be considered 
a two dimensional layered system.

\section{Methodology}

We have taken the experimental structure of NaTiO$_2$ in its high temperature
phase from Clarke et al. \cite{clarke}. We have performed a detail electronic 
structure calculation for the rhombohedral phase of NaTiO$_2$ using standard 
full potential linearized augmented plane wave method which is implemented in 
WIEN2k code\cite{wien2k}.
For the experimental structure, the muffin tin sphere radii were chosen to be
2.23, 2.07 and 1.83 a.u. for Na, Ti and O respectively. Plane wave cutoff
parameter R$_{mt}$K$_{max}$ was choosen as 7.00 and approximately 120 
k-points were used over the irreducible first Brillouin zone.
Convergence has been achieved to energy values less than 1 meV 
per formula unit with respect to the variation of number of k-points and the parameter RK$_{max}$. 

As NaTiO$_2$ contains a transition metal ion having one d electron in the 
outermost shell, Coulomb correlations are expected to be significant in this 
system. Therefore, we have performed calculations within local spin density 
approximation (LSDA) and the LSDA+U approximation (which includes orbital Coulomb correlation). 
More than one implementation is available for the latter
method generating different double counting corrections for the underlying LSDA functional.
 For moderately correlated (or metallic) systems a mean field
approximation to the d-orbital part of the LSDA functional is considered to be 
appropriate. This is described in details as the Around Mean Field
(AMF) functional in ref.\cite{amf}. As NaTiO$_2$ is observed to be a bad 
metal in both high and low temperature phase experimentally\cite{clarke}, 
we have chosen the AMF method in our LSDA+U calculations reported below. 
In this work we consider a range of U values (3.3 - 4 eV) with emphasis on two in particular: 3.6eV and 4.0eV.
The former is obtained theoretically via constrained LDA calculations by the authors of ref.\cite{ezhov}, specifically for this compound. This value is very close to that used (U=3.3eV) for compounds like TiOCl, NaTiSi2O6, (La,Y)TiO3 which gives good agreement with experimental data like magnetic susceptibility\cite{seidel}, spin gap \cite{streltsov}, optical measurements \cite{lmtolapw}, respectively. 
Therefore we take it as a physically reasonable value. The latter is considered merely for theoretical purposes.   

\section{Results and Discussion}

\subsection{Experimental structure: ground state}
\label{elecstr}

\begin{figure}
\begin{center}
\includegraphics[width=6cm]{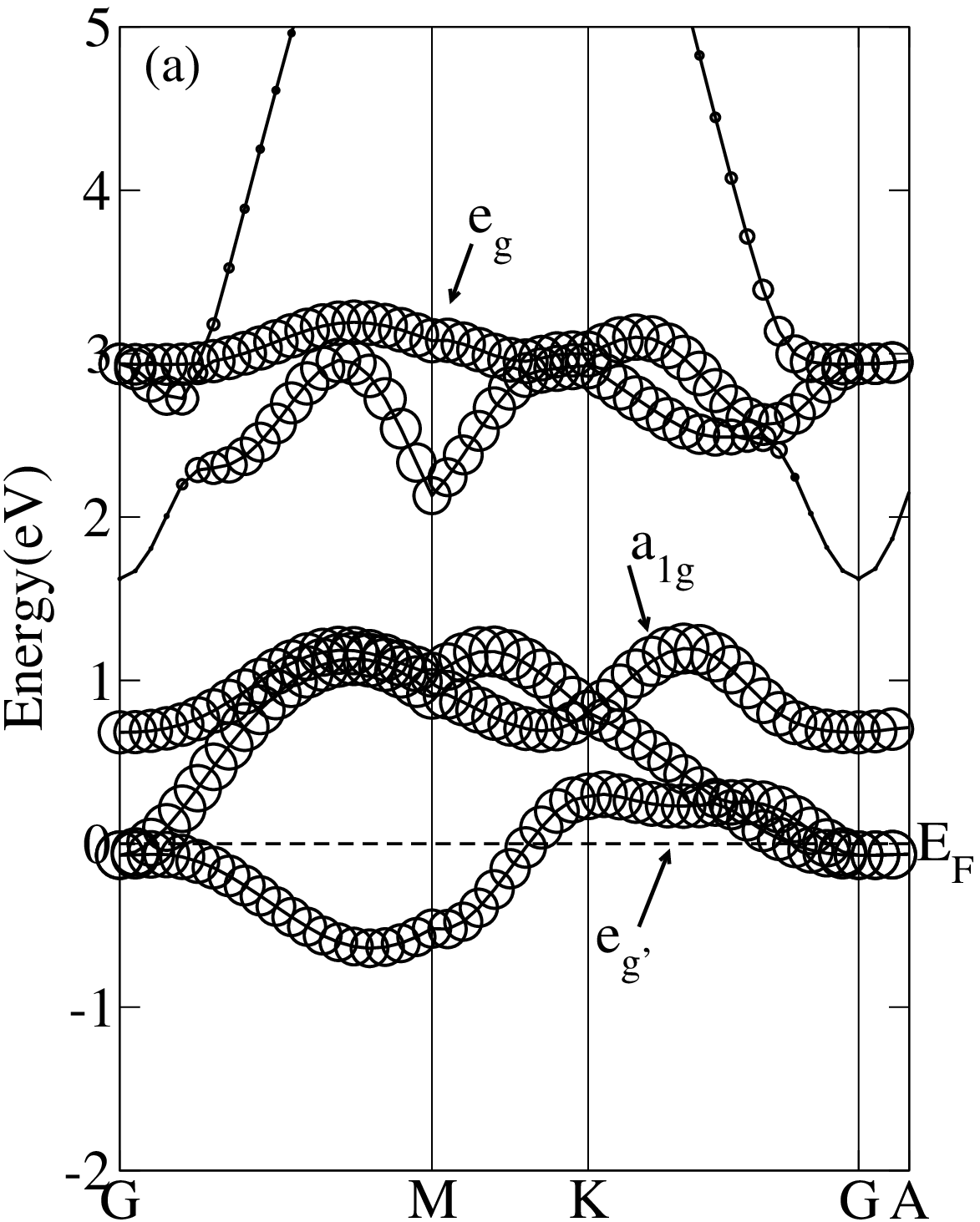}\includegraphics[width=6cm]{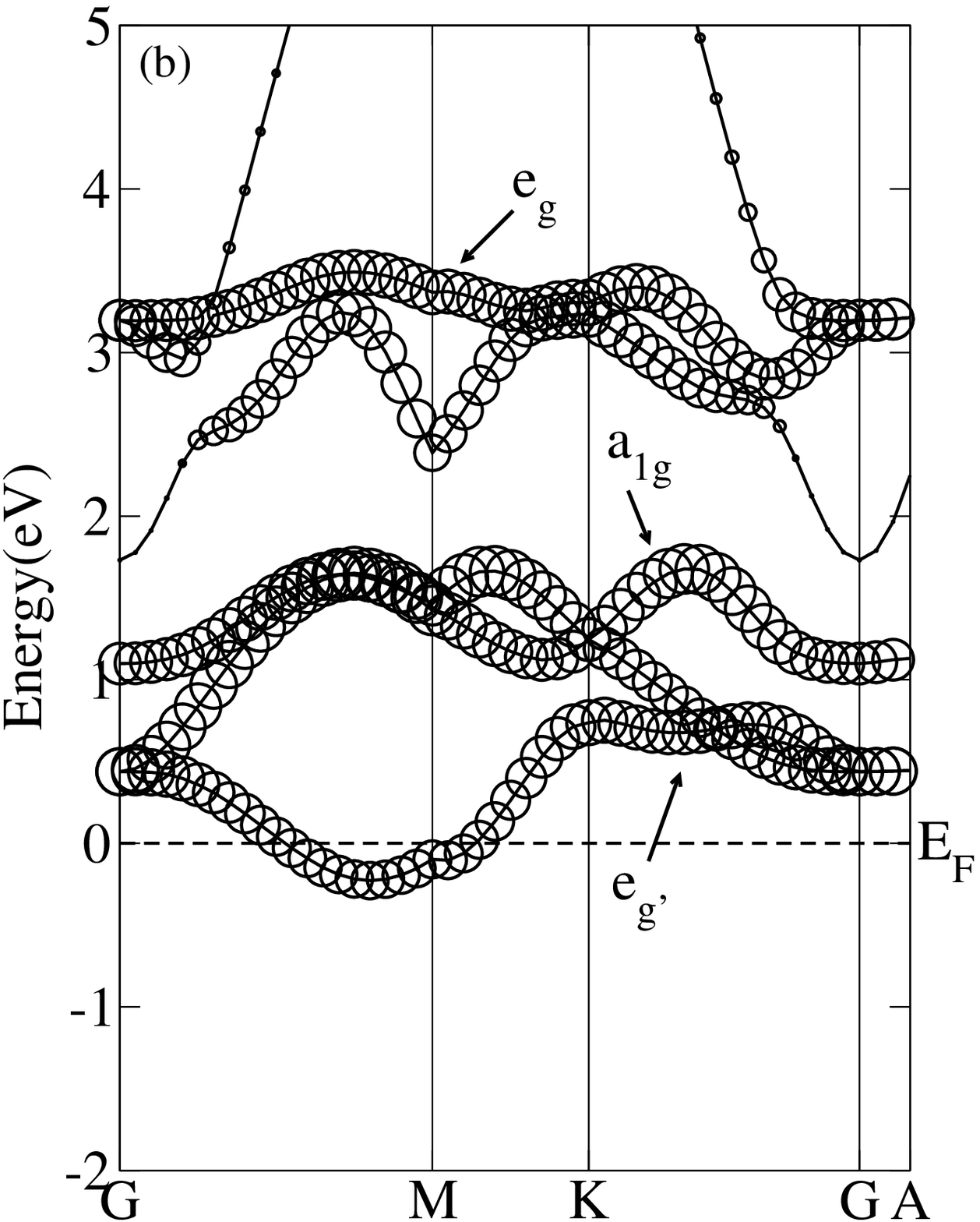}
\caption{Band structure in both (a) spin up  and (b) spin down channel 
         for experimental structure within LSDA.}
\label{bslsdaexp}
\end{center}
\end{figure}

We first study the electronic ground state of the experimental crystal structure within the LSDA and LSDA+U functionals. 
Fig.~\ref{bslsdaexp} shows the electronic band structure within LSDA.
The ground state is found to be metallic with Ti d states 
present at the Fermi level in both spin directions. 
One also observes that after the octahedral crystal field splitting only t$_{2g}$ states lie at the Fermi level. These are then further split due to the trigonal crystal field of oxygens into two sub-manifolds: \egp and \ag, the former being doubly degenerate.

\begin{figure}
\begin{center}
\includegraphics[width=6cm]{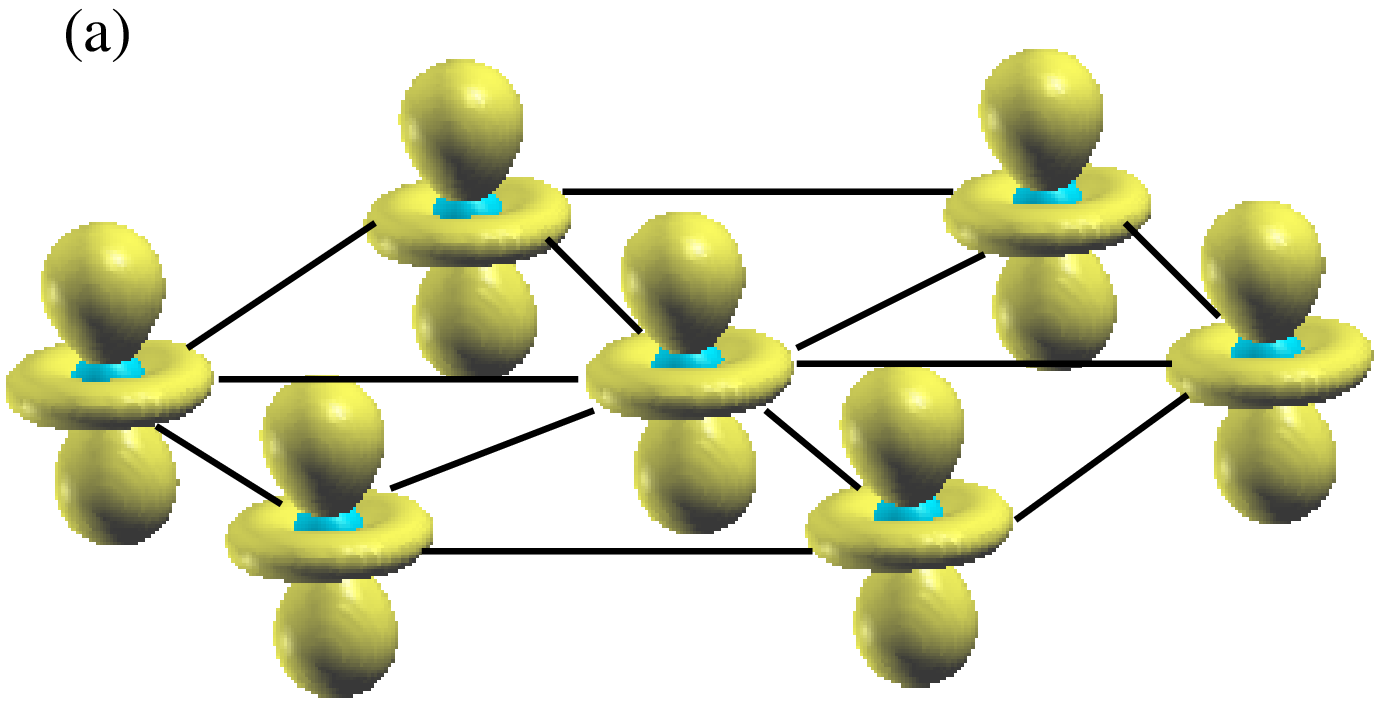}\includegraphics[width=2cm]{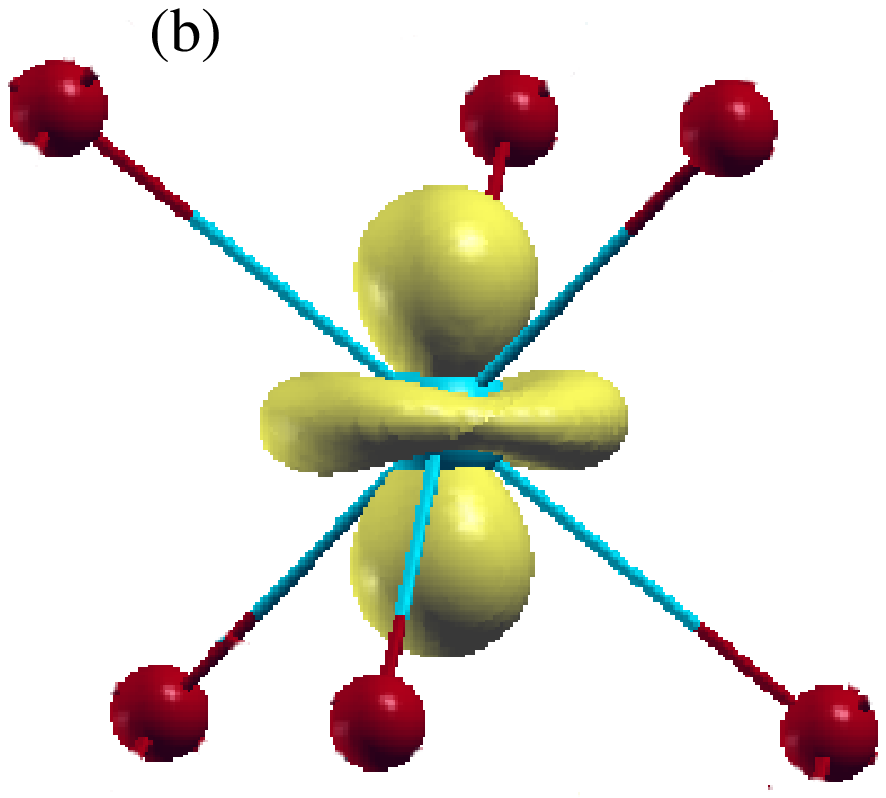}
\caption{The real space electron density (a) in the Ti plane showing 
the \ag orbital at each site and (b) within the oxygen octahedron.}
\label{density-0.23}
\end{center}
\end{figure}

In the system $xyz$ of Fig.~\ref{crysstr} these orbitals have the following representations (see ref. \cite{terakura}) based on canonical d-orbitals, 
\begin{eqnarray*}
\mbox{\ag}      &:& 3z^2-r^2 \\
\mbox{\egp}     &:& 1/\sqrt 3  [ yz+ \sqrt 2 xy ] \\
& & 1/\sqrt 3  [ zx - \sqrt 2 (x^2-y^2) ]
\end{eqnarray*}
The \ag orbital has the typical $d_{z^2}$ shape along the trigonal axis and perpendicular to the Ti plane (see Fig.~\ref{density-0.23});
as a result the \ag-\ag direct overlap of $\sigma$ type is very small. Direct overlap is mainly of $\pi$ type generating a small bandwidth ($\sim$0.6eV) as evident from Fig.~\ref{bslsdaexp}. 
The electron density in real space with \ag orbital occupied at all Ti sites is shown in Fig.~\ref{density-0.23}. The negligible \ag-\ag direct overlap of $\sigma$ type is clearly evident from Fig.~\ref{density-0.23}. 

The \egp orbitals are doubly degenerate and have 4 different contributions from the canonical d orbitals. 
The $xy$ and $x^2-y^2$ components have maximum amplitude within the $xy$ plane ($z=0$) and generate an almost circularly symmetric density providing distributed overlap over all six coordinating Ti atoms. Their direct overlap is of $\sigma$ type (therefore much stronger than the \ag-\ag overlap) and produces an efficiently delocalising wave function. This is reflected in Fig.~\ref{bslsdaexp}  by the bandwidth of the \egp bands of almost 2 eV.
The $yz$ component has all 4 lobes $\pi$-bonding along the $y$ axis where the Ti-Ti distance is shortest. 
The $xz$ component has only two lobes remaining from its linear combination with $x^2-y^2$ within the $xz$ plane: the two pointing towards the oxygen (direction $z= \sqrt 2 x$) vanish and the two pointing in between the other two oxygens are conserved.

From band structure analysis we see that the bands cutting the Fermi level have predominantly \egp character. 
In particular the $yz$ component is present in the lowest band and the $xy$ and $x^2-y^2$ components in the next one. 
The highest band of the $t_{2g}$ group has mainly \ag character.

In Table~\ref{occnr} we present electron population analysis within LSDA i.e. the occupation numbers of the $t_{2g}$ bands up to the Fermi level (excluding the valence band contributions due to bonding). 
As mentioned above, we note that the occupied bands (labelled \egp in Fig.~\ref{bslsdaexp}) have a non neligible character of type \ag. 
This is due to the significant trigonal distortion of the oxygen octahedron which causes the mixing of the two $t_{2g}$ subgroups.
The LSDA functional however predicts a slight predominance of \egp over \ag occupation (for comparison the former should be divided by two due to its degeneracy). This is explained by the fact that \egp states allow the hopping of electrons over a dense 2d triangular network of Ti atoms (high coordination number and effective orbital overlap in contrast to analogous compound like f.i. TiOCl~\cite{tiocl}) 
thus generating a significant energy gain due to electron delocalisation.  
Since they are doubly degenerate and not fully occupied (1 electron per Ti 
atom) the electronic structure is metallic.

\begin{table}
\begin{center}
\begin{tabular}{|c|c|c|c|c|}
\hline
        &  \ag    & \egp     \\
\hline
spin up & 0.17 & 0.40  \\
spin dn & 0.05 & 0.08  \\
\hline
\end{tabular}
\end{center}
\caption{LSDA orbital occupation numbers of the trigonal projections within t$_{2g}$ manifold. }
\label{occnr}
\end{table}
  
\begin{figure}
\begin{center}
\includegraphics[width=6cm]{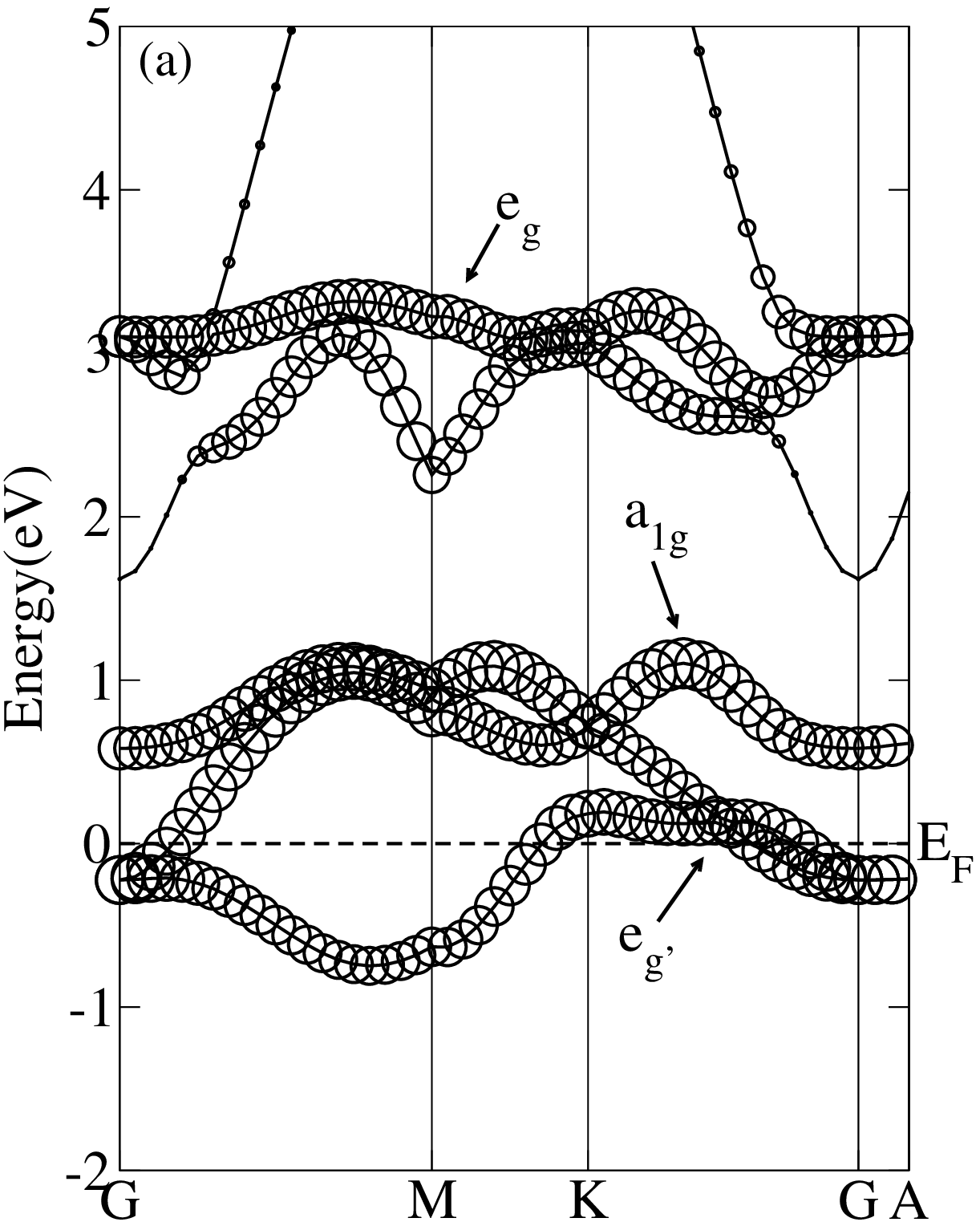}\includegraphics[width=6cm]{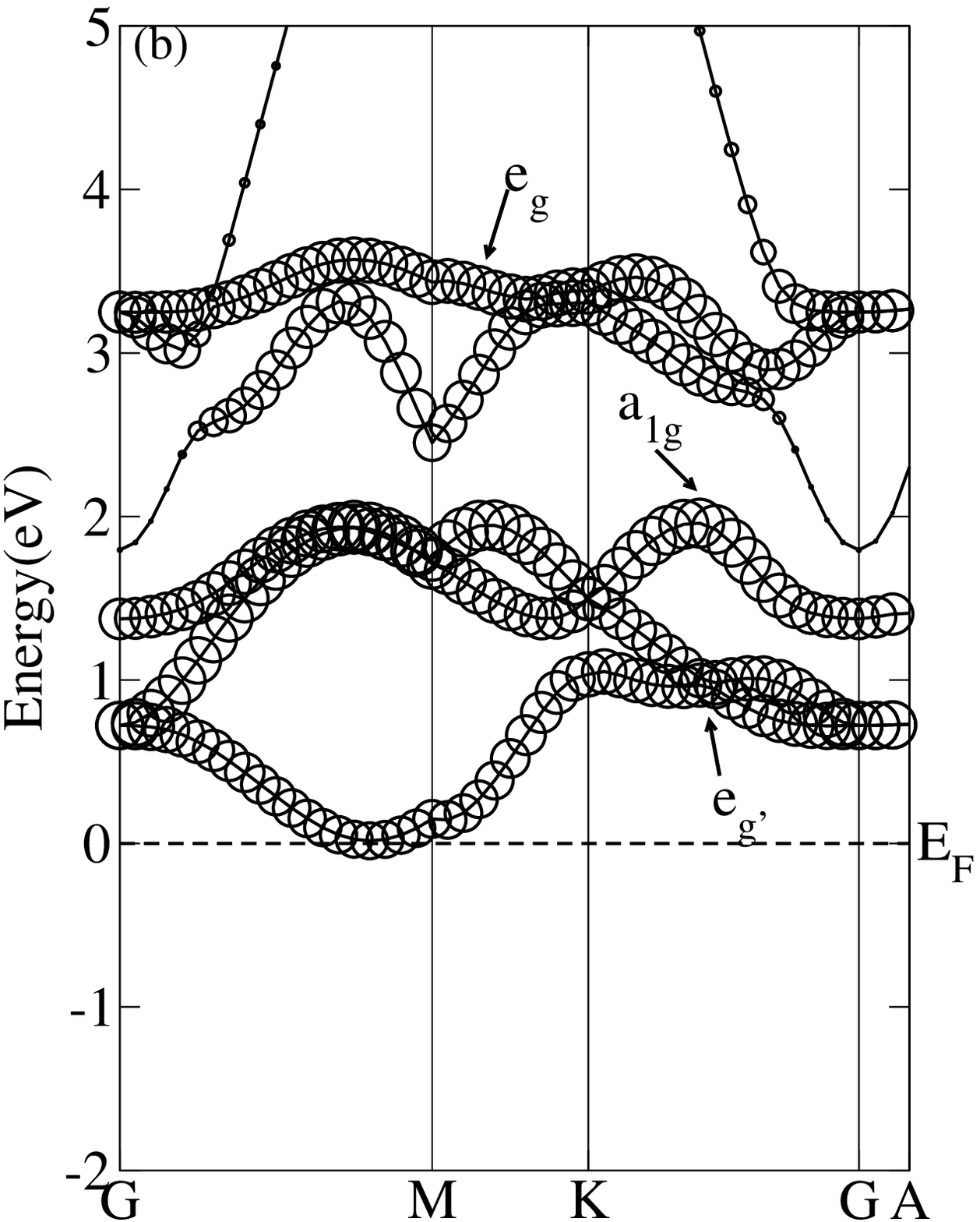}
\caption{Band structure in (a) spin up and (b) spin down channels within LSDA+U with U = 3.6 eV.}
\label{bslsdauexp}
\end{center}
\end{figure}

For transition metal atoms like Titanium the effect of local electron correlation U is analysed via the LSDA+U functional.  Fig.~\ref{bslsdauexp} shows the band structure in both majority and minority spin channels for a physically reasonable value of U = 3.6 eV for Titanium\cite{ezhov}. 
The effect here consists merely in a complete spin polarisation of the bands. Due to the degeneracy of the \egp states at the Fermi level, no further orbital polarization could be induced by the orbital Coulomb correlation $U$ within LSDA+U approximation. Therefore, the two majority spin bands remain half filled and the system behaves as a metal.

\begin{figure}
\begin{center}
\includegraphics[width=6cm]{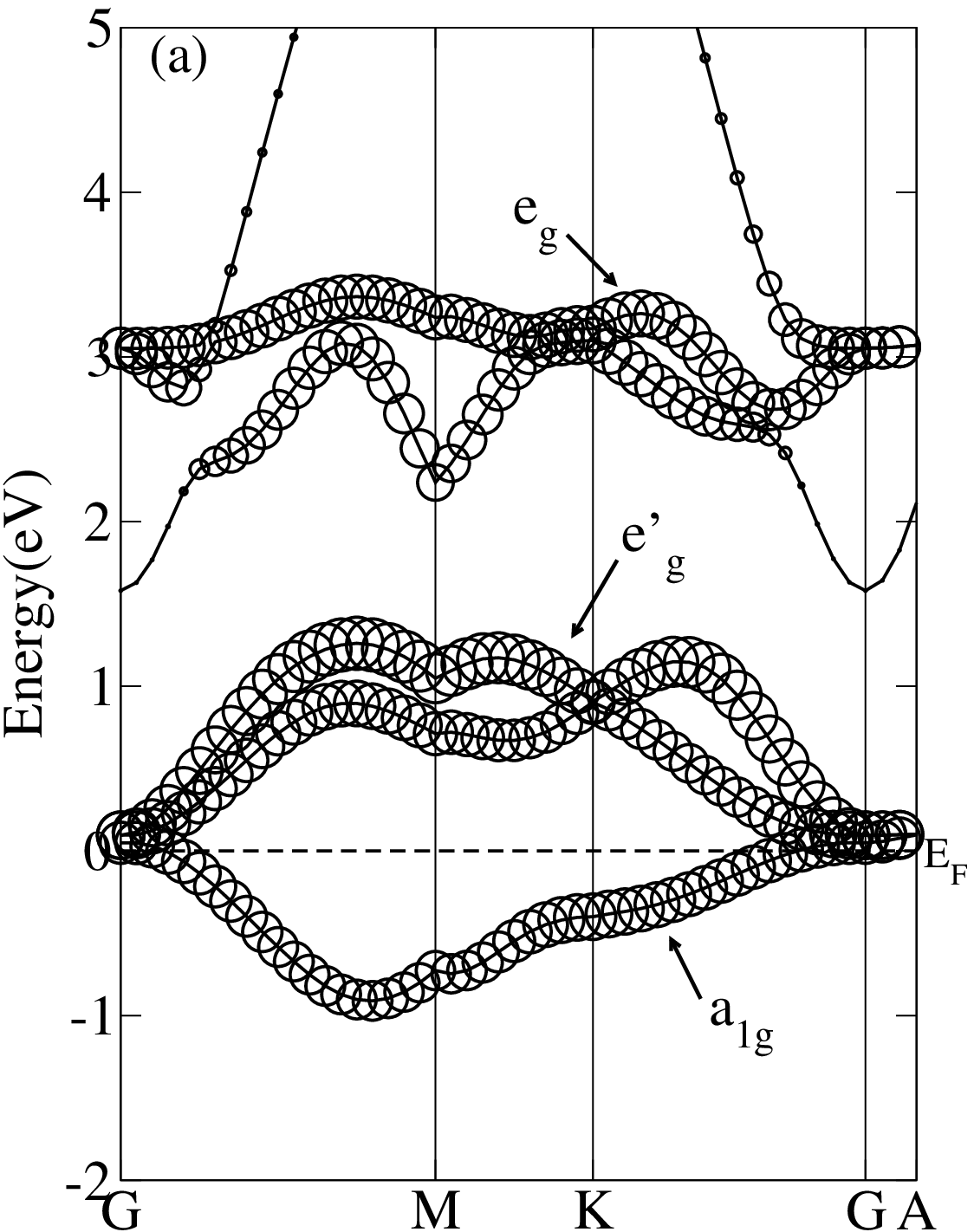}\includegraphics[width=6cm]{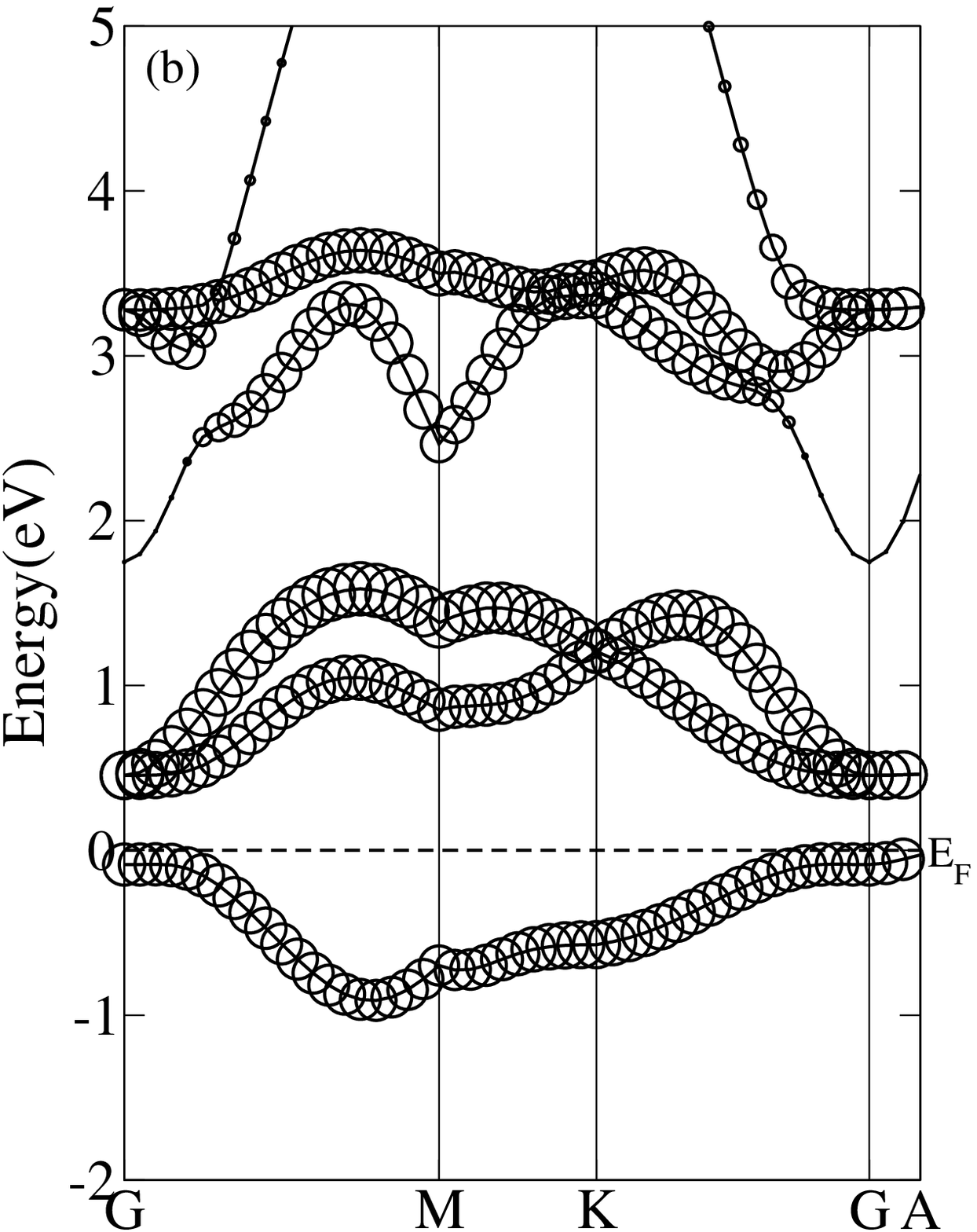}
\caption{Band structure for ES1 state (see text) in majority spin channel within LSDA+U with (a) U=3.6 eV and (b) U = 4 eV.}
\label{bslsdauagexp}
\end{center}
\end{figure}

From Table~\ref{occnr} we note that the occupation number of the \ag orbital
is smaller but rather close to the \egp one (divided by two due to its degeneracy). This suggests that the electronic excitation with \ag orbital occupied is a low energy one. In order to estimate the magnitude of this excitation energy, we took the approach of orbital locking. In this approach, the occupation of a chosen orbital is kept fixed till convergence and then relaxed within a second convergency run. The orbital occupation is fixed by manually setting the initial density matrices (see appendix for details). We call the electronic configuration with \ag orbital occupied predominantly as ES1 state and that with \egp orbital occupied predominantly as ES2 state, hereafter. Note that for U=3.6eV ES2 is the ground state.
By forcing the single Ti d-electron to occupy the \ag orbital and then allowing it to relax (ES1) we obtain a stable solution with a total energy about 5meV (per formula unit) higher than the ground state one (Fig.~\ref{bslsdauagexp}), using U=3.6eV. This state represents a collective excited state and its electronic structure is illustrated in Fig.~\ref{bslsdauagexp} (a). It shows the feature of an almost zero gap semiconductor therefore having extremely limited conductivity. 
  
Increasing U to a theoretical value of 4 eV, the ES1 state becomes the ground state whereas ES2 is now an excited state by a tiny energy difference (see next discussion on dependence on U). Fig.~\ref{bslsdauagexp} (b) shows the ES1 state with a single \ag band below the Fermi level and the opening of a gap in the majority spin channel due to a stronger U. In comparison to previous U value, here the conductivity is totally absent. 

\begin{figure}
\begin{center}
\includegraphics[width=9cm]{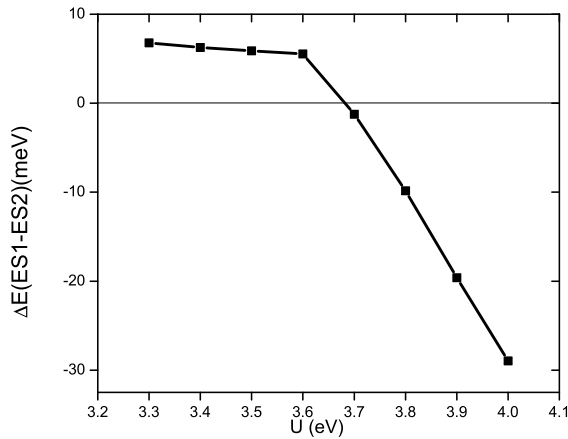}
\caption{Total energy difference between ES1 and ES2 states with varying U}
\label{dEvsU}
\end{center}
\end{figure}

Based on this result and given the proximity in energies of the ES1 and ES2 
states we study their stability as DFT solutions with varying U.
This is obtained using the approach of orbital locking as mentioned earlier. 
The plot of the resulting total energy difference of the two solutions is presented in Fig.~\ref{dEvsU}.
We see that for U values in the region considered physically reasonable for 
Titanium (around 3.6eV) the ES2 state is the ground state but lower in energy from the ES1 state by only about or less than 5 meV (this is well above the error bar mentioned in the methodology section). As U crosses the value 3.7 eV, ES1 becomes the ground state with the gradual opening of a gap. For U=4eV the gap magnitude in the majority spin channel amounts to about 0.4 eV (see Fig.~\ref{bslsdauagexp}(b)). 

The above results show the interplay between two standard and competing mechanisms (energy scales)
in transition metal compounds: the localisation (U) favoring \ag occupation and the delocalisation (kinetic energy) favouring \egp. 
This holds for a wide range of temperatures up to room temperature but for very low temperature ($<$ 5mev $\simeq$ 50 K) the excited state cannot be thermally reached and only the \egp state is populated.

For the physical value of U=3.6eV our results are inline with experimental results by Clarke~\cite{clarke} whereby a bad metal with low but non zero conductivity is found. The bad metallic behaviour is ascribed to the existence of an almost insulating state (ES1,\ag) very close in energy (5mev) to a conducting ground state (ES2,\egp).

Previous calculations on the high temperature phase of this system was 
reported in reference~\cite{ezhov} where the authors found an insulating 
behaviour within the LSDA+U (U=3.6eV) approximation with a spin-polarised 
a$_{1g}$ band being fully occupied and a gap of about 1 eV. 
Though we observe the insulating ES1 state as ground state for higher U values (U $>$ 3.7eV), for U=3.6eV we clearly see that ES2 state is the ground state. More importantly, we observe that these two electronic states (ES1 and ES2) have very closeby energies over a reasonably large range of U values (see Fig. 6). \\
We would also like to mention that in the calculation reported in reference~\cite{ezhov}, already within LSDA the \ag occupation was obtained
to be higher than that of \egp which is opposite to what we observe here (see table 1).  We believe that this discrepancy is due to the 
adoption of the Linear Muffin Tin Orbital (LMTO) basis set in the previous 
calculations contrary to Linearized Augmented Plane Wave (LAPW) basis used here. LMTO is known to be deficient in describing the interstitial region in 
between atomic (muffin tin) spheres with respect to the LAPW method
(see sec. III of ref.~\cite{lmtolapw} as an example).
The appropriate description of this region is fundamental 
to take into account electron delocalization effects via a realistic overlap of
the orbitals of \egp symmetry. 
Finally, ref.~\cite{ezhov} finds the compound to be insulating in contrast 
to experimental results~\cite{clarke}. Hence the results presented in this work
better describes the experimental situation.

\subsection{Lattice effects: optimization of oxygen position}

\begin{figure}
\begin{center}
\includegraphics[width=7cm]{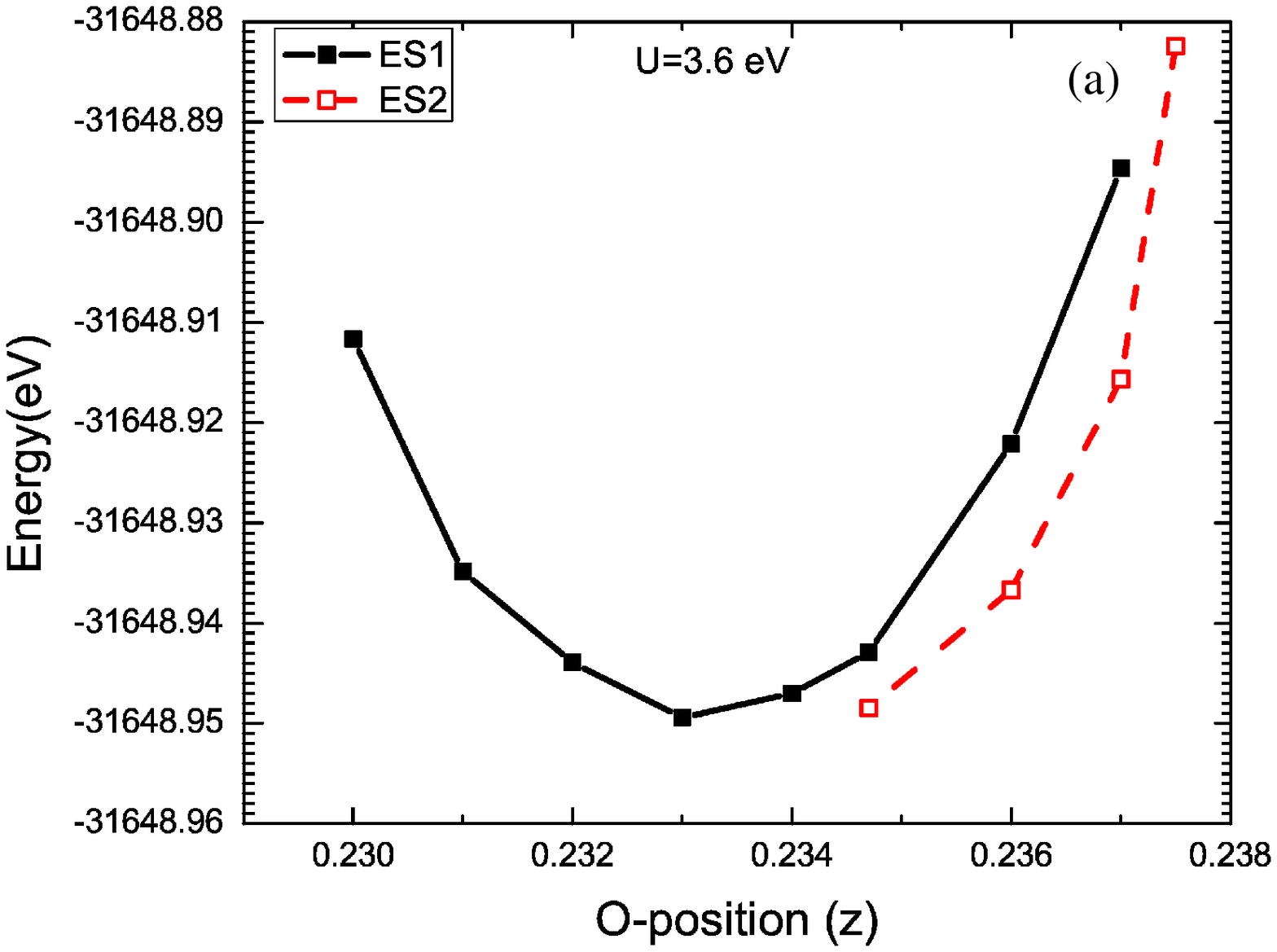}
\includegraphics[width=7cm]{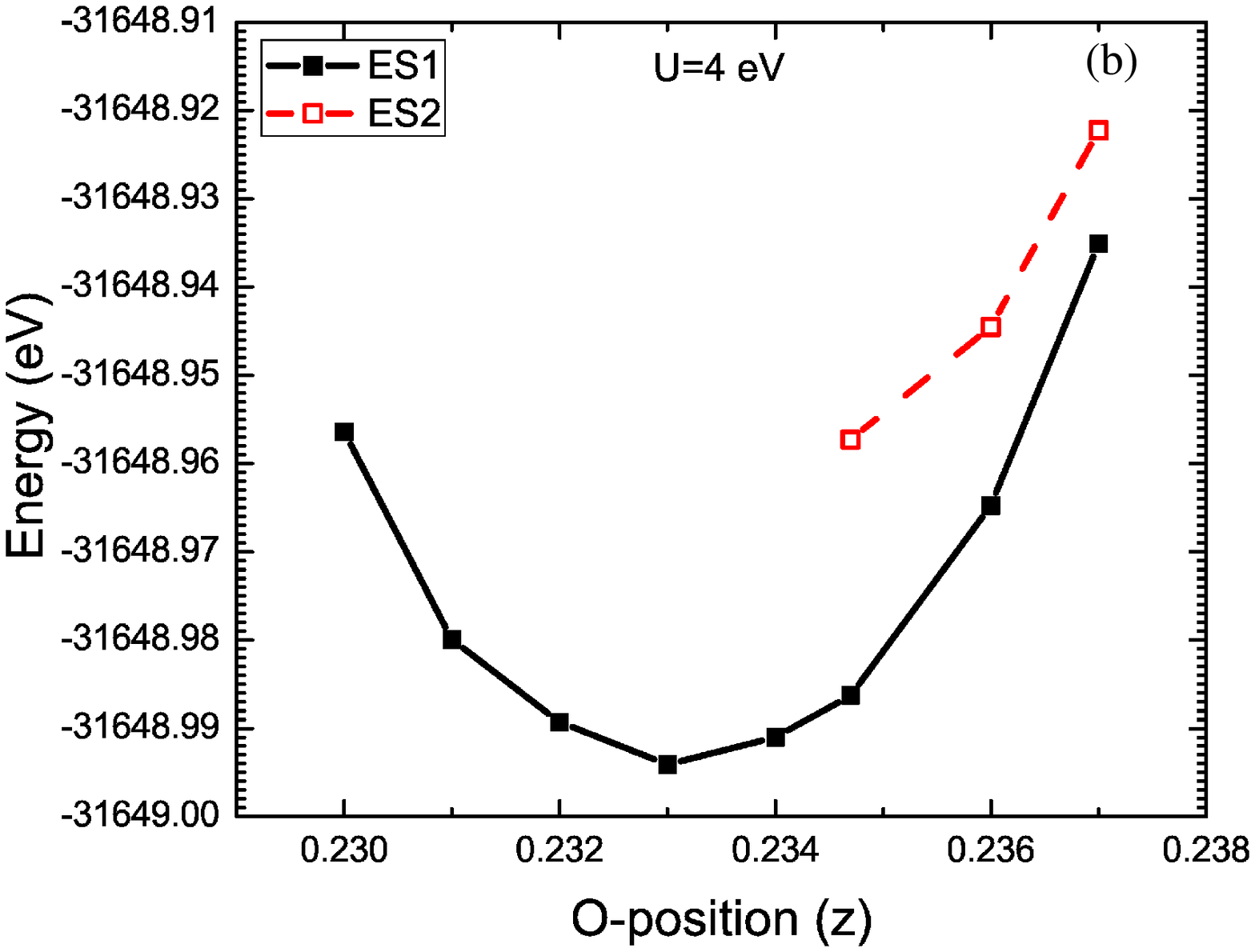}
\caption{Total energy as function of oxygen coordinate for ES1 (full curve and squares) and ES2 (dashed curve and empty squares) states for (a) U=3.6eV and (b) 4eV.}
\label{crossover}
\end{center}
\end{figure}

Following an ab-initio approach we investigate the equilibrium structure  
within the high temperature phase symmetry which allows
only for oxygen displacements along the z-axis (Fig.~\ref{crysstr}). 
Given the low dimensionality of the problem, we choose to optimize the oxygen 
positions using the total energy minimization method which is based on drawing 
the energy curve as a function of the oxygen coordinate (z,z,z)~\cite{tiocl}.
Lattice parameters are kept at their experimental values. This has the advantage
of simulating the effect of van der Waal's forces between layers given that they are not taken into account within DFT. 
As learnt in previous section, the system presents two almost degenerate 
groundstates for relevant U values: ES2 (ground state) and ES1 (low lying excited state). 
Therefore we study the stability of both states as a function of the oxygen coordinate applying 
the orbital locking method as in previous section. 
\\
In Fig.~\ref{crossover} we present the results for two values of U.
The full curve represents the total energy of ES1 state and the dashed one is 
that of ES2 state. 
For the theoretical U (=4eV) (fig. 7b) the ES1 state is much lower in energy than the ES2 one and represents the groundstate for all oxygen positions.
(Note that the ES2 state is not stable for all oxygen positions, as indicated by the dashed curve having no value below the experimental z (=0.2348). This is probably due to the crystal field of oxygen atoms preventing the electron delocalisation within ES2 state.)
However, for the physical U (3.6eV) (fig. 7a) the energies of ES1 and ES2 become comparable for several oxygen positions. As a consequence there exists a range of oxygen positions (0.233,0.235) for which the total energy is almost constant. As a result oxygen thermal vibrations would be characterised by large amplitudes and generate pronounced orbital fluctuations between \ag and \egp states.

\begin{figure}
\begin{center}
\includegraphics[width=10cm]{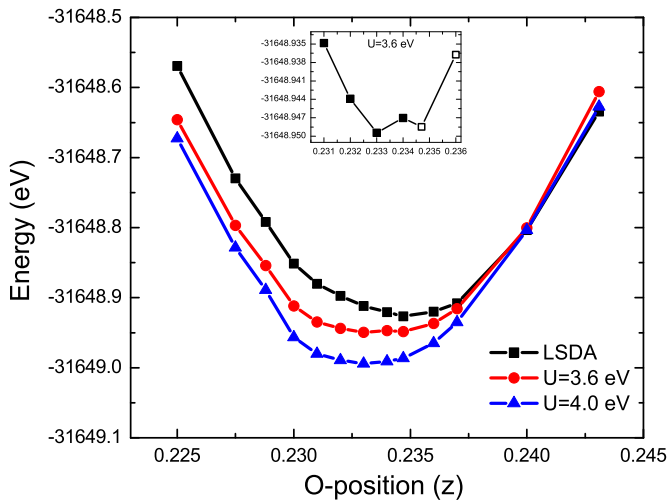}
\caption{Total energy versus oxygen positions within LSDA and LSDA+U.
Oxygen position z is given in fractional coordinate. 
z$_{exp}$=0.2348 corresponds to the experimental structure. 
Smaller values of z represent shorter Ti-O bond length. 
Inset shows a closer view of the energy landscape near the equilibrium point for U=3.6 eV; empty symbols represent ES2 state and full ones ES1 state.} 
\label{optOxyg}
\end{center}
\end{figure}

The overall energy landscape for three values of U is shown in Fig.~\ref{optOxyg}. Within the LSDA functional (U=0), the experimental structure is found to be the equilibrium one. On the contrary for the physical U (3.6eV), the energy landscape is clearly flat as it is for U=4eV but to a lesser degree. In the inset, a zoom-in of the energy curve for U=3.6eV is given 
showing quantitatively the flatness of the energy curve: its variation remains below 20 mev in a wide range of z values (0.231, 0.236).
At room temperature this landscape gives rise to soft oxygen vibrational modes with quite large oscillation amplitudes which enhance orbital fluctuations significantly.

It is worth mentioning that the optimization of lattice parameters does not alter the conclusions 
drawn above qualitatively. For example, optimization within LDA leads to reduction in both
$a$ and $c$ parameters by about 2$\%$. The reduction of $a$ causes a greater overlap of \egp orbitals 
thus making ES2 more stable. Competing to this is the shrinking of $c$ which favours ES1 (\ag) due
to the crystal field of oxygen being closer to \egp orbitals. 

We note that the above scenario suggest a possible explanation for the anomalous value of the specific heat jump at the structural transition: this is in fact found to be bigger than the one for pure spin system by Takeda {\it et. al}~\cite{takeda}. Clarke {\it et. al} \cite{clarke} have extracted from Takeda's measurements a value of the entropy change of 7 J K$^{-1}$ mol$^{-1}$.
In case of long range antiferromagnetic ordering of S=1/2 spins the entropy change should be R ln(2S + 1) = R ln 2 = 5.8 J K$^{-1}$ mol$^{-1}$\cite{clarke}. No magnetic long range order is found in this system and short range antiferromagnetic correlations could only give rise to an entropy change
lower than R ln(2S + 1). This would be even further away from the experimental value of 7 J K$^{-1}$ mol$^{-1}$.
In this work we have shown at length the relevance and interplay of orbital degrees of freedom. As there are both experimental and theoretical predictions\cite{cava, dhariwal} for long range orbital order (with partial orbital polarization at each site) at low temperatures, the entropy change is expected to be lower but indicatively close to R $ln(2*|Orb| + 1)$ where the quantum number $|Orb|=1$ represents the three orbitals: \ag and \egp. This upper bound amounts to 9.1 J K$^{-1}$ mol$^{-1}$ which would make the entropy change due to orbital order with partial orbital polarization close to the experimental 7 J K$^{-1}$ mol$^{-1}$.

\section{Conclusions}

We have performed a detail electronic structure calculation for the high temperature rhombohedral phase of NaTiO$_2$. 
We first study the electronic properties of the (non optimised) experimental structure. The ground state is found to be conducting with \egp orbital symmetry. At an energy slightly greater than the ground state we find an excited state with \ag orbital symmetry and almost fully  insulating. By thermal excitation the latter is populated comparably to the ground state giving rise to an overall low conductivity. This is inline with what found by Clarke et al. \cite{clarke} in their experiments.  Secondly, we study the variation of the above properties with respect to lattice via static distortions of oxygen positions along the trigonal axis. Depending on the oxygen proximity to the Ti plane
the \ag (oxygens close to the Ti plane) or \egp (oxygens distant from the Ti plane) orbitals will be alternately populated. Moreover we find the energy landscape to be rather flat which gives rise to a large vibrational amplitude. This implies that at room temperature the population of \ag or \egp orbitals will oscillate significantly.
The specific heat jump at 250 K detected by Takeda et al.\cite{takeda} could, therefore, be tentatively accounted for by an entropy contribution coming from orbital degree of freedom rather than spin.

\section{Acknowledgement} 

This work is supported by CSIR, India project grant 
(ref. no. 03(1212)/12/EMR-II). TM acknowledges I. Singh and A. Taraphder for useful discussions.  
\section{Appendix}
The initial density matrices ($5\times 5$) supplied for the ES1 and ES2 electronic state within orbital lock approach are in the basis (l=-2,-1,0,1,2) as described below. For ES1 state the diagonal elements
of the spin up density matrix are set to (0.0, 0.0,1.0,0.0,0.0) and the remaining matrix elements to zero. 
Therefore only the d$_{z^2}$  (l=0) or \ag orbital is populated. The spin down density matrix is identically zero. 
For ES2 state we have set the diagonal elements of the spin up density matrix to (0.25, 0.25,0.0,0.25,0.25) 
and the remaining matrix elements to zero. Here we considered the simplest occupation spectrum with the \ag orbital empty: 
this consists in populating uniformly all the remaining harmonics (l= -2, -1, 1, 2) therefore both the e$_g$ and \egp orbitals,
counting onto the fact that once the orbital lock is removed self consistency will surely depopulate the e$_g$ orbitals due to their energetics.

\end{document}